\documentclass[fleqn,usenatbib]{mnras}

\usepackage[T1]{fontenc}

\DeclareRobustCommand{\VAN}[3]{#2}
\let\VANthebibliography\thebibliography
\def\thebibliography{\DeclareRobustCommand{\VAN}[3]{##3}\VANthebibliography}

\usepackage{graphicx}	
\usepackage{amsmath}	
\usepackage{amssymb}	

\usepackage{newtxtext,newtxmath}

\def\apj{ApJ}
\def\apjl{ApJL}
\def\aap{A\& A}

\def\araa{ARA\&A}
\def\mnras{MNRAS}

\def\aj{{AJ}}
\def\apjs{ApJS}

\def\mnras{{MNRAS}}

\def\pasp{{PASP}}

\def\alamenos#1{$^{-#1}$}

\newcommand{\cmtres}{\rm cm^{-3}}

\newcommand{\EF}{{\cal EF}}    %

\newcommand{\Egrav}{E_{\rm grav}}    %
\newcommand{\Ekin}{E_{\rm kin}}    %

\newcommand{\kms}{\mathrm{km~s^{-1}}}

\newcommand{\Msun}{$M_\odot$}    %

\newcommand{\SFRff}{{\rm SFR_{\rm ff}}}

\newcommand{\tauff}{\tau_{\rm ff}}

\newcommand{\tcross}{t_{\rm cross}}

\newcommand{\trelax}{t_{\rm relax}}

\title[Violent relaxation]{{\textcolor{black}{Gravity or turbulence V: Star forming regions undergoing violent relaxation.}} 
}

\author[Andrea Bonilla-Barroso, J. Ballesteros-Paredes et al. ]{Andrea Bonilla-Barroso$^{1}$, Javier Ballesteros-Paredes$^1$, Jesus Hernández$^2$,
\newauthor Luis Aguilar$^{2}$, Manuel Zamora$^4$, Lee W. Hartmann$^5$,    Aleksandra Kuznetsova$^6$
\newauthor  Vianey Camacho$^{1,4}$, Verónica Lora$^{1,3}$,
\thanks{E-mail: a.bonilla@irya.unam.mx}
\thanks{NHFP Sagan Postdoctoral Fellow}
\\
\\
$^{1}$Instituto de Radioastronom\'ia y Astrof\'isica, UNAM, campus Morelia. PO Box 3-72. 58090. Morelia, Michoac\'an, M\'exico
\\
$^{2}$Universidad Nacional Aut\'onoma de M\'exico, Instituto de Astronom\'ia, AP 106,  Ensenada 22800, BC, México
\\
$^{3}$Instituto de Ciencias Nucleares, Universidad Nacional Aut\'onoma de M\'exico, Apartado Postal 70-543, 04510, M\'exico City, M\'exico
\\
$^{4}$Instituto Nacional de Astrof{\'i}sica, {\'O}ptica y Electr{\'o}nica, Luis E. Erro 1, 72840 Tonantzintla, Puebla, M{\'e}xico
\\
$^{5}$Dept. of Astronomy, University of Michigan, 1085 S. University Ave, Ann Arbor, MI 48109, USA
\\
$^{6}$American Museum of Natural History. Central Park West at 79th Street. 10024. New York, NY. USA.
}

\date{Accepted 2022 January 26. Received 2022 January 25; in original form 2021 November 16}

\pubyear{2022}

\begin{document}
\label{firstpage}
\pagerange{\pageref{firstpage}--\pageref{lastpage}}
\maketitle

\begin{abstract}

{
Using numerical simulations of the formation and evolution of stellar clusters within molecular clouds, we show that \textcolor{black}{the stars in clusters} formed within collapsing molecular cloud clumps exhibit a constant velocity dispersion \textcolor{black}{regardless of their mass, as expected in a violent relaxation processes}. In contrast, clusters formed in turbulence-dominated environments exhibit an {\it inverse} mass segregated velocity dispersion, where massive stars exhibit larger velocity dispersions than low-mass cores, consistent with massive stars formed in massive clumps, which in turn, are formed through strong shocks. We furthermore use Gaia EDR3 to show that the stars in the Orion Nebula Cluster exhibit a constant velocity dispersion as a function of mass, suggesting that it has been formed by collapse within one free-fall time of its parental cloud, rather than in a turbulence-dominated environment during many free-fall times of a supported cloud. 
{Additionally, we have addressed several of the criticisms of models of collapsing star forming regions}: namely, the age spread of the ONC, the comparison of the ages of the stars to the free-fall time of the gas that formed it, the star formation efficiency, and the mass densities of clouds vs the mass densities of stellar clusters, showing that observational and numerical data are consistent with clusters forming in clouds undergoing a process of global, hierarchical and chaotic collapse, rather than been supported by turbulence.
}
\end{abstract}




\section{Introduction}\label{sec:intro}

It is widely accepted that most of the stars form in clusters \citep{Lada_Lada03}, 
and most of these clusters become loose associations that expand throughout the Galaxy \citep{Gouliermis18}. However, the very process of cluster formation is complicated. New stars are created and interact with each other, while also are subject to the global gravitational potential of the model cloud, which furthermore, varies in time as it collapses, first, and dissipates away, later.

There are two competing scenarios of how stars form. On one hand, the turbulent scenario (TS), in which molecular clouds (MCs) are supported against collapse by supersonic (and/or super-Alfvénic) turbulence \citep[see reviews by, e.g.,][and references therein]{Vazquez-Semadeni+00, MacLow_Klessen04, Ballesteros-Paredes+07, McKee_Ostriker07, Klessen_Glover16}. In this scenario, clouds are assumed to be in virial equilibrium between self-gravity, turbulence, (sometimes) magnetic fields and (frequently) confined by external pressures \citep[][]{Myers_Goodman88a, Myers_Goodman88b}. In such environment, stars form along the many free-fall times that the cloud lasts, at a low rate, keeping low the efficiency of the whole cloud \citep[e.g., ][]{Krumholz+19}.  

Much of this interpretation relies on the fact that the \citet{Larson81} scaling relations for MCs appear to imply that clouds are turbulent, and their energies are in virial balance. Neither of these assumptions, however, are necessarily true \citep{Ballesteros-Paredes+06, Ballesteros-Paredes+20}. For instance, regarding the first Larson relation, the non-thermal velocity dispersion interpreted as supersonic turbulence could very well be the result of collapse \citep{VS+07, Ballesteros-Paredes+11a}. In addition, clouds exhibit a variety of energies with a large scatter which, furthermore, could be wrongly estimated by gross assumptions due to the observational limitations and definitions \citep{Ballesteros-Paredes+18}. Regarding the  second Larson relation, the constancy of the column density of clouds has been found to be an artifact consequence of our definition of clouds and the low filling factor of the dense structures \citep{Ballesteros-Paredes+12, Beaumont+12, Ballesteros-Paredes+20}.

On the other hand, the global, hierarchical and chaotic collapse scenario \citep[GHCC, see ][and references therein]{Vazquez-Semadeni+19}, which proposes that non-uniform, irregular molecular clouds are necessarily in a permanent state of global contraction towards local centers of collapse \citep{Vazquez-Semadeni+07, Heitsch_Hartmann08, Ballesteros-Paredes+11a}. In this scenario, molecular clouds are born from large-scale \rm HI flows\footnote{The very origin of such flows may be diverse: from turbulent flows in the ISM, expanding shells due to previous events of star formation, or large-scale gravitational (Toomre, Parker, etc.) instabilities, see \citet[][and references therein]{Dobbs+14}}, and a variety of (magneto-)hydrodynamical instabilities \citep{Heitsch+05, Heitsch+06, Heitsch+07a, Heitsch+07b} along with thermal instability \citep{Hennebelle_Perault99, Audit_Hennebelle05}, allow to amplify density fluctuations, producing the rich inner structure observed in MCs \citep[see, e.g., ][]{Vazquez-Semadeni+19}. Having many Jeans masses, clouds necessarily tend to collapse globally, but since the timescales for collapse are faster for denser regions, these are able to rapidly collapse and form massive stars, which in turn destroy their parent cloud. In this scenario, the observed relation by \citet[][]{Heyer+09} between the Larson's ratio, ${\cal{L}}=\sigma_\upsilon/R^{1/2}$ (with $\sigma_\upsilon$ the velocity dispersion, and $R$ the size of the cloud), and the column density $\Sigma$, is the natural outcome of the process of collapse: once gravity dominates and drives inward motions in a hierarchical and chaotic way, clouds' energies tend to a virial-type equipartition between the kinetic ($\Ekin$) and gravitational ($\Egrav$) energies \citep{Vazquez-Semadeni+07, Ballesteros-Paredes+11a, Ballesteros-Paredes+18}. In a sense, thus, \citet{Heyer+09} relation is nothing but a generalization of the \citet{Larson81} relation for non-constant column densities. 

It should be stressed that this virial-type equipartition does not imply that clouds are in virial equilibrium, in the sense that their moment of inertia $I$ has second time derivative equals zero, $\ddot{I}=0$. What the results of \citet{VS+07} and \citet{Ballesteros-Paredes+11a} show is that a collapse situation naturally produces $\Egrav\sim 2\Ekin$ after $\sim$ one free-fall time, even though these are non-equilibrium situations,
a result that also occurs for pure $N-$body systems \citep{Noriega-Mendoza_Aguilar18}.

In order to distinguish between the different models of star formation, it may be useful to look at the actual {kinematics of the stars in young clusters}. 
%

\textcolor{black}{While kinematics of stars in young clusters may help distinguish between the turbulent vs. collapsing pictures,
what should be observed in any particular situation depends upon a number of factors.
For example, one might expect that collapse models straightforwardly predict the stars have systematic inward motions.  However, numerical simulations show that once a cluster is forming, the motions of the stars are very quickly randomized,
so the signature of collapse is wiped out except perhaps on the distant outskirts 
\citep[see Figures 6 and 7 of ][ and section \S\ref{subsec:kinematics} for a more quantitative picture]{Kuznetsova+15}.  At later times, 
it has been argued that the collapse model will always produce expansion motions of the stars after gas exhaustion or clearance, while the turbulent model should exhibit always random motions \citep[{e.g.,}][]{Ward+20}.
This however is not necessarily a distinguishing factor.
The amount of expansion, and thus the resultant velocities, will depend upon the efficiency of
star formation as well as the timescale over which the gas is dispersed, and whether the escaping stars vs. the remnant bound cluster
dominate the observations \citep{Mathieu1983, Lada+84}.
In fact, it is not obvious that a turbulent model, in which the stars are virialized with the gravitational potential of both stars and gas, would not also expand if the gas represents a significant fraction of the gravitating mass and is lost quickly.}

\textcolor{black}{We envisage two ways to distinguish between the two models. On one hand, if star formation occurs over a significant time interval, the collapse model would predict that the older stars are more spatially spread, and the younger stars would be in a more compact configuration because they form later when the gas has collapsed further; while the turbulent support model would predict that the young stars should have the same spatial distribution as the older stars.}

A different useful approach could be to group the stars in mass bins, and look at the velocity dispersion of each mass bin. At first glance, a cluster that is collisionally relaxed (a situation that could be thought to occur in turbulent clouds) should exhibit energy equipartition between their stars: more massive stars must exhibit a velocity dispersion smaller than low-mass stars, at a proportion given by energy equipartition:
\begin{equation}
  \bigg( \frac{\sigma_{\rm high}}{\sigma_{\rm low}} \bigg)^2 
  = 
  \frac{M_{\rm low}}{M_{\rm high}}
  \label{eq:equipartition}
\end{equation}
where $\sigma_{\rm high}$ is the velocity dispersion of massive stars, $\sigma_{\rm low}$ is the velocity dispersion of  low-mass stars, and $M_{\rm high}$ and $M_{\rm low}$ are   their masses, respectively. This will be valid if the age of the cluster is larger than its collisional relaxation time, defined as 
%
\begin{equation}
  t_{\rm relax} \sim \bigg(\frac{0.1 N_*}{\ln{N_*}}\bigg)\ t_{\rm cross}
  \label{eq:trelax}
\end{equation}
%
with $N_*$ the number of stars in the cluster \citep{Binney_Tremaine08}. For $N_*\sim$1000$-$2000, as is the case of the Orion Nebula Cluster, ONC \citep{DaRio+12, Robberto+20}, this quantity is between 14 and 26 times larger than the crossing time, typically larger than the age estimation of the cluster. Thus, young clusters are hardly relaxed.

\begin{figure}
\centering
{\setlength{\fboxsep}{0pt}%
\setlength{\fboxrule}{1pt}%
\fbox{\includegraphics[width=0.95\columnwidth]{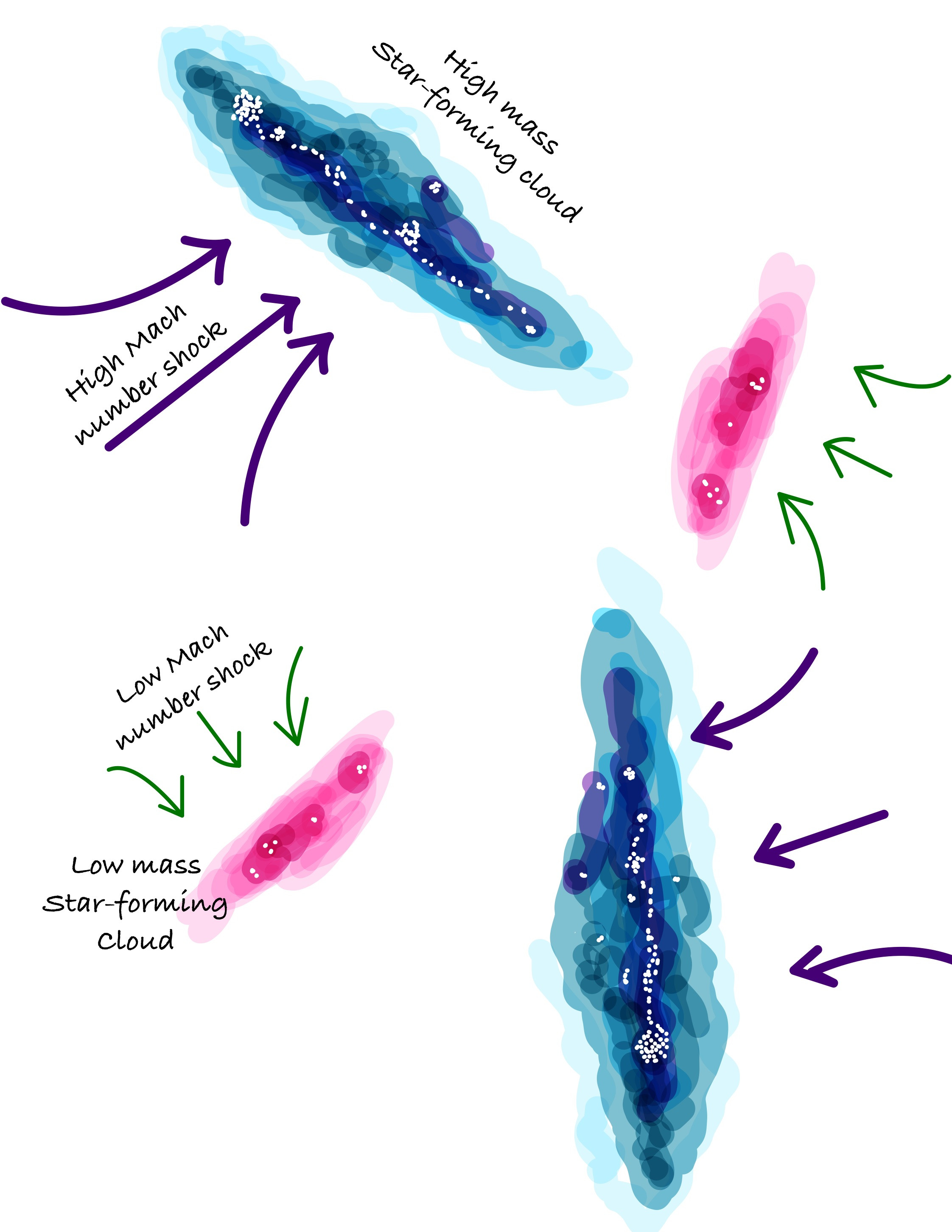}}}
\caption{Schematic diagram of clouds formed by turbulent compressions. In this scheme, larger clumps and cores are formed by larger shocks, and thus, one can expect more massive clumps and cores to exhibit larger velocity dispersions than low-mass cores.}
\label{fig:esquema_veldis_turb}
\end{figure}

Contrary to the belief that clusters born in turbulent environments should be collisionally relaxed, massive stars could very well have a larger velocity dispersion compared to low-mass stars. A simple mechanism may be at play. First, that massive stars are formed within massive cores, while low-mass stars are formed in, statistically speaking, lower-mass cores. In a turbulent cloud, massive cores are formed by stronger shocks compared to low-mass cores. Then, one can expect that the core-to-core velocity dispersion of massive cores is larger than the velocity dispersion of low-mass cores. Since the stars inherit the velocity of their parental core, one should also expect that the velocity dispersion of massive stars to be larger than that of the low mass stars. This situation is schematically depicted in Fig.~\ref{fig:esquema_veldis_turb}, which is meant to be the interior of a molecular cloud. In this figure, we draw massive cores formed by larger turbulent compressions with blue, and low-mass cores, formed by smaller compressions, with pink. As can be seen graphically, in a turbulence-dominated environment, massive cores require stronger shocks, compared to low-mass cores, and thus, the core-to-core velocity dispersion within the cloud will be larger for massive cores than to low-mass cores. Thus, massive stars, which are born from massive cores, will tend to have larger a velocity dispersion than low-mass stars, formed in cores with lower masses.

Such mechanism is not unique, and another one may be at play: be gravitational heating of high-mass stars. Since gravity has a negative heat capacity, massive stars tend to eject low-mass stars from the center of clusters and associations, loosing energy, sinking into more bound orbits, and becoming kinetically hotter. This mechanism makes equipartition impossible, as described by \citet{Spitzer69}, and shown in recent $n-$body numerical simulations \citep{Parker+16, Spera+16, Webb_Vesperini17}.
{Whether one or the other effect dominates in turbulent clouds\footnote{A point of caution is in order: global collapse models are not exempt of the \citet{Spitzer69} instability. It may very well be at play as soon as the gravitational potential slows down its variation, as is suggested by the results of \citet{Parker+16}.} will be investigated elsewhere (Bonilla-Barroso et al., in preparation). The relevant point is that one can expect that turbulent clouds will form massive stars that will exhibit larger velocity dispersions than low mass stars. }



By contrast, in the case of a global, hierarchical collapse model, the resulting velocity dispersion of stars will be independent of their mass, as changes to their velocities during violent relaxation are due to the overall change in the potential rather than gravitational interactions between stars
\citep{Lynden-Bell67}. It should be noticed furthermore, that in this case, even if the stars within the clusters randomize their orbits during the formation of the cluster, the velocity dispersion per mass bin of one or the other cases are expected to stand during the first stages of the life of the cluster, as long as the gravitational potential keeps changing due to the collapse of the cloud, and/or the relaxing time is longer than the lifetime of the cluster. 


In the present work we analysed the velocity structure of a suite of numerical simulations meant to model either collapsing and turbulent MCs, under different configurations, as well as the velocity structure of the ONC, using Gaia EDR3 data.  In \S\ref{sec:methods}  we present either the numerical simulations and the observational data. The results of our analysis is presented in \S\ref{sec:results}, and extensively discussed in \S\ref{sec:discussion}. 

\section{Methods and Data}\label{sec:methods}




In order to be able to interpret the observational data through the ONC, it becomes necessary to first look at the velocity dispersion of stars and cores in numerical simulations, for which we know the actual physical stage, and we can compute observational statistics. We notice that, in what follows, for each field in the simulated and observational data, we group the stars under analysis in mass bins with the same number of objects, in order to have similar statistics

\subsection{Numerical simulations}\label{sec:simulations}

In the following sections we analyse six different numerical simulations that represent the  evolution of a molecular cloud. The first two where performed with GADGET-2 \citep{Springel05}, which is a smoothed particle hydrodynamics (SPH) code, with the inclusion of sink particles as proposed by \citet{Jappsen_Klessen05}. The remaining four were performed using Flash~4.3 \citep{Fryxell+00}, with the inclusion of sink particles as proposed by \citet{Federrath+10}. These simulations represent different physical conditions of molecular clouds, according to the current models of cloud formation and evolution. We briefly describe these simulations, quoting the corresponding papers for details of each run.

\subsubsection{Global collapse I: collapse of a super-Jeans cloud with decaying supersonic turbulence.}

\noindent {\bf Run 1.} Taken from \citet{Ballesteros-Paredes+15}, this simulation represents the interior of a 1~pc box with 1,000~\Msun, a constant density field, and a supersonic (Mach 8) initial turbulent velocity field that is left to evolve under the action of self-gravity over $\sim$ one free-fall time. 

 \noindent {\bf Run~2} was presented by \citet{Kuznetsova+15}. The simulation is similar to Run~1, but has an initial gass mass of 2300~\Msun, a density field which has a semi-ellipsoidal shape, with dimensions $2\times3\times4$~pc, and a smooth density gradient towards the vertex of the ellipse. The velocity field has the superposition of a turbulent field with a velocity dispersion with Mach 8, and an initial rotation \citep[see][for details]{Kuznetsova+15}.


\subsubsection{Global collapse II: collapsing cloud, originally build up from colliding diffuse gas streams.}


\noindent {\bf Run 3.} \textcolor{black}{Presented by \citet{Zamora-Aviles+18}, describes the formation and evolution of a molecular cloud from the diffuse warm atomic medium under typical Solar Neighborhood conditions \citep[see, e.g., ][]{VS+07, Heitsch_Hartmann08}.
The numerical box, of sizes $L_{x,y,z}=20 \, {\rm pc}$ is initially filled with warm neutral gas at uniform density of $10~\cmtres$ and constant temperature of $160~K$. This temperature corresponds to the thermal equilibrium and implies an isothermal sound speed of $c_{\rm s}\simeq3.1\kms$. We consider open boundary conditions in the $y-$ and $z-$directions and turbulent inflows (with a subsonic turbulent Mach number of 0.7) entering in the $x-$direction at a velocity of $5 \, \kms$, the typical velocity dispersion of the HI in the Solar Neighborhood. }

We include the physical processes relevant to study the formation and evolution of molecular clouds, such as magnetic fields (ideal MHD), heating and cooling, self-gravity, and sink formation. For heating and cooling we use the analytic fits by \citet{KI02} as implemented by \citet{Vazquez-Semadeni+07}. 
Sink particles can be formed when the density in a given cell exceeds a threshold number density, $n_0 \simeq 10^{6}~\cmtres$, it is bound, and the region exhibits local converging motions (i.e., negative divergence), among other standard sink-formation tests as discussed by \citet{Federrath+10}. The numerical box is initially permeated by a uniform magnetic field of $3$~$\mu$G along the $x$-direction, a value consistent with the observed mean value magnetic field of the uniform component in the Galaxy \citep{Beck01}. The resolution in the center of the numerical box, where the cloud forms, is uniform ($1024^3$) and decreases two levels of refinement towards the $x-$edges, where locally the resolution is 256$^3$.


\subsubsection{Clouds with supersonic turbulence}

\noindent {\bf Runs 4, 5 and 6} are three simulations of continuously driven isothermal turbulence. The numerical box size is 1 pc per side and contains 1,000 \Msun. It is discretized in a regular grid of $512^{3}$. In all cases, the injected kinetic energy is a mixture of {50$\%$ solenoidal and 50$\%$}\ compressive supersonic turbulence, with Mach numbers of 5, 10 and 15 respectively. It is injected at large scales with a power spectrum of $E(k) \propto k^{-2}$, where $k$ is the wavenumber which is in the range of 1$-$3 . The gas in each simulation is continuously forced in two steps. First, without self-gravity during $\sim$5 crossing times. After this time the turbulence is statistically homogeneous, and thus, we allow for sink formation by turning-on self-gravity during 20$\%$ of the free-fall time, in order to have enough sinks, but avoiding as much as possible the influence of self-gravity. The sink formation recipe is the same as run 4, but with a critical density of $\simeq 5.5 \times 10^7$cm$^{-3}$. 

Since we did not evolve these simulations much during the period of sink creation, a single realization of Runs 4, 5 and 6 will create only a relatively small number of sink particles. To increase the statistics, thus, these simulations were repeated 3 times each one, with different random seeds.



%
%
%

\subsection{Empirical sample}\label{sec:observations}

We have derived stellar masses and ages for a sample of kinematic candidates selected using the astrometric observables of parallax ($\varpi$) and proper motions ($\mu_\alpha$, $\mu_\delta$) from GAIA-EDR3 \citep{GAIAEDR3} in the chosen studied region (hereafter, the ONC region: $83.0^\circ\le\alpha_{J2000}\le84.5~\circ$ and $-6.5^\circ\le\delta_{J2000}\le-4.0^\circ$). We apply the selection criteria used by Hernandez et. al. (in preparation), which we describe here for clarity: 

%
\begin{enumerate}
  \item We first apply the zero-point correction of parallax following the method described by \citet{Lindegren2021a}. 

  \item We restrict the astrometric quality of the sample, requiring a  Re-normalised Unit Weight Error (RUWE)$<$1.4 \citep{Lindegren2021b} and a parallax error smaller than 20\%.
  

%

  \item We then selected stars with parallaxes in the range $2.27< \varpi <2.93 \; \rm mas$ and stars with a proper motion modulus ($\mu=\sqrt{\mu_\alpha^2+\mu_\delta^2}$) 
  $<3.23  \; \rm mas \, yr^{-1}$. Based on the $\varpi$ and $\mu$ distributions of the kinematic young members selected by \citet{Kounkel2018} in the the ONC region, these limits were defined using the median and the standard deviation applying a 3$\sigma$ criteria. 
 \end{enumerate}

With these filters, we obtain a sample of 3030 kinematic candidates.
We first locate these stars with high precision in the H-R diagram, and use models of pre-main-sequence stellar evolution, in order to infer their masses and ages.
For this purpose, we cross-match our kinematic candidates with the optical spectroscopic census of Hernandez et al. ({hereafter the optical spectroscopic sample}; in preparation), finding spectral types for 1586 stars. Then, we used the {\tt MassAge} code (Hernandez et al., in preparation) to compute ages and masses of 1461 stars. The sample without masses and ages includes the star 2MASS\_J05343988-0625140, which falls below the main sequence, and 124 stars which do not have reliable $J-$band magnitude.\\

In brief, the {\tt MassAge} code uses spectral types (or effective temperatures), photometry ($G_p$, $R_p$, and $B_p$) and parallaxes from GAIA-EDR3 \citep{GAIAEDR3}, and the $J$ and $H$ magnitudes from 2MASS \citep{Cutri2003}. The uncertainties in the estimated values are obtained using the Monte Carlo method of error propagation \citep{Anderson1976}. The extinction is obtained by comparing the observed colors with the standard colors reported in \citet{Luhman2020}. Here, we use the extinction law from \citet{CCM1989} to obtain the extinction in each photometric band normalized at 0.55{\micron} ($A_\lambda/A_V$). The luminosity is estimated from the $J-$band absolute magnitude using the bolometric correction from \citet{PM2013}.  We convert spectral types into effective temperatures using the standard table of \citet{PM2013}. Finally, we obtain stellar masses and ages by comparing the location of the stars on the Hertzsprung–Russell diagram with theoretical values. Here, we use two evolutionary models: the PARSEC model \citep{Marigo2017} and the MIST model \citep{Dotter2016}. 

\begin{figure}
\includegraphics[width=\columnwidth]{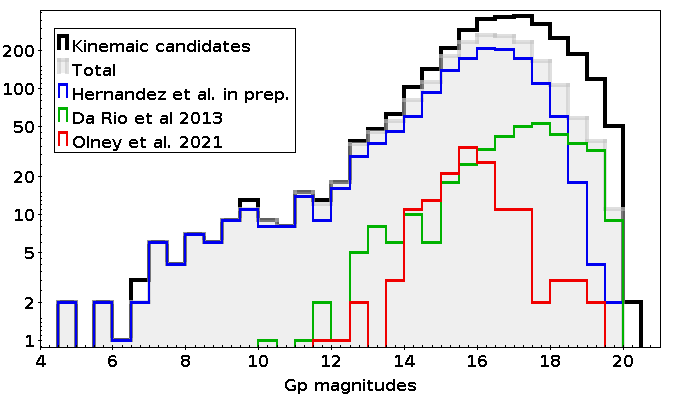}
\caption{Distribution of the Gp GAIA-EDR3 magnitudes of the kinematic candidates, including stars with derived stellar masses. Blue histogram includes the sample studied by Hernandez et al. (in preparation). Green and red histograms represent the additional stars from \citet{DaRio2012} and \citet{Olney2020}, respectively. The filled grey histogram represents the 1989 stars with derived masses and ages. }
\label{fig:Gban-dist}
\end{figure}

In order to expand our sample, we have included 382 and 146 additional stars studied by \citet{DaRio2012} and \citet{Olney2020}, respectively. \citet{DaRio2012} reported effective temperatures using two narrow-band filters located at 7530 {\AA} and  7700   {\AA} (tracing the continuum and a TiO-band feature, respectively) and a broad I-band filter, taking into account the extinction law from \citet{CCM1989}. 
Additionally, using a deep convolutional neural network, the APOGEE-net tool \citep{Olney2020} improves the effective temperature and the stellar surface gravity derived by \citet{Kounkel2018} by comparing APOGEE H-band spectra (R $\sim$22500) and PHOENIX spectral theoretical library \citep{Husser2013}. 
On the other hand, the optical spectroscopic sample includes 429 and 697 stars studied by  \citet{DaRio2012} and \citet{Olney2020}, respectively. Generally, the effective temperatures derived in those works agree within 500K with the effective temperatures derived in the optical spectroscopic sample.
Using the effective temperature as input parameter in the MassAge code, we derived stellar masses and ages for stars studied by \citet{DaRio2012} and \citet{Olney2020}, not included in the optical spectroscopic census of Hernandez et al. (in preparation).  Figure \ref{fig:Gban-dist} indicates that the additional stars increase the number of faint kinematic candidates ($G>12$) with estimated masses (likely low mass stars). We include a total of 1989 stars with ages and masses (filled histogram), representing more than 65\% of the kinematic candidates.

\begin{figure}
\includegraphics[width=\columnwidth]{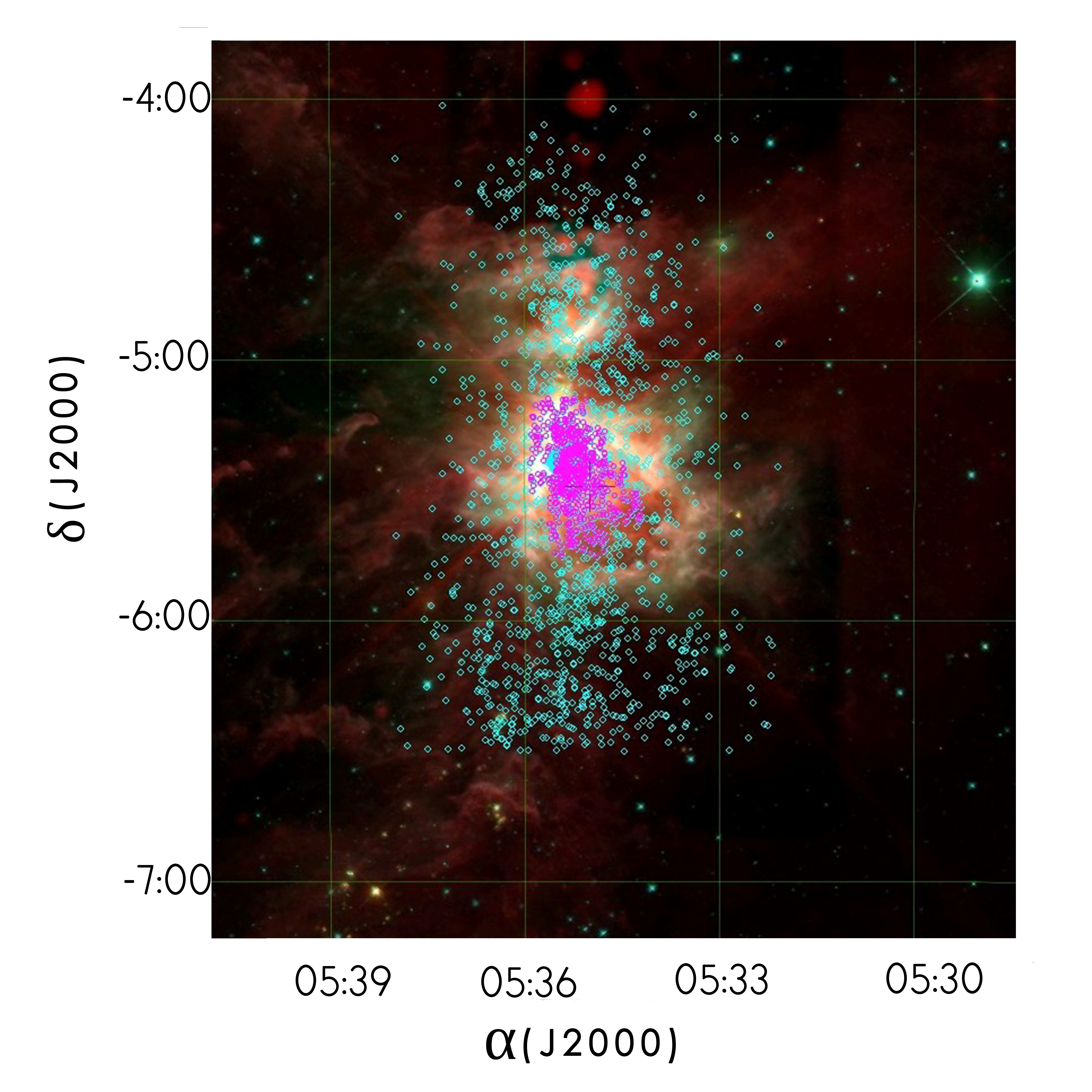}
\caption{Image taken by Wide-Field Infrared Survey Explorer (WISE) showing the spatial distribution of the ONC used in this analysis. The cyan and magenta points represent the central and extended sample, respectively.}
\label{fig:ONCmap}
\end{figure}

These sources are spread over an area of 7$\times$7~pc, which we call the extended area, or the total sample. In addition, we distinguished the central core, a region of about 1.5$\times$1.5~pc, determined as the overdensity having more than 30 stars per square degree, 3 times the rms value of the extended area. Fig.~\ref{fig:ONCmap} shows a map of the analysed region. The stars in the extended region are shown as blue circles, and the stars in the central cluster as magenta circles.

\section{Results}\label{sec:results}

{
The main goal of the present contribution is to analyse the velocity dispersion of the sinks per mass bin either in our numerical simulations of molecular clouds, as well as in the stars of the ONC. However, since for the ONC we had to infer the ages and masses of the stars, it is worth mentioning first some points regarding these derived distributions.

\subsection{Ages and masses of the stars in the ONC}

Fig.~\ref{fig:AgeHistogramONC} shows the age histograms of our two samples of the ONC. The top panel shows the ages inferred using MIST models, while the lower panels show the corresponding distributions using PARSEC models.  Some points are worth stressing. First of all, that each model produces a statistically different age distribution from the other model. For the total sample (blue histograms), for instance, the MIST models produce younger distributions than the PARSEC models, while for the the central sample (red histograms), the situation is reversed and the MIST models are the ones that produce older distributions.  


Second, the number of stars older than 5~Myr strongly decreases in the case of the MIST models. In comparison, the PARSEC models have still important contribution of stars for bins up to 10~Myr. We will discuss this and the previous point in \S\ref{sec:AgeSpreadONC}.

Third, we notice that the central sample (red histograms) is statistically younger than the extended or total sample (blue histograms) regardless the evolutionary model, a fact known to occur also in other clusters \citep[e.g., ][]{Getman+14, Getman+18}.

Fourth, in the present work we will consider only the masses and proper motions of the stars younger than 10~Myr in each sample, in order to minimize contamination by older dispersed populations and/or possible stars with edge-on disks that appear older because of disk obscuration. We notice that the contribution of older than 10~Myr stars to the statistics shown in the next sections will be negligible:  in the case of the total sample, we are rejecting only 5\%\ (MIST) and 13.7\%\ (PARSEC) of the stars, while in the case of the central core, 2\%\ (MIST) and 9\%\ (PARSEC). 

\begin{figure}
\includegraphics[width=\columnwidth]{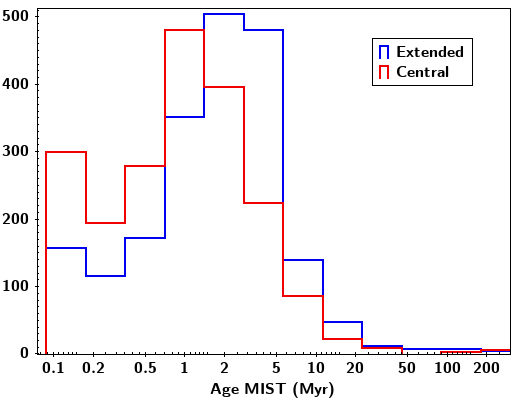}
\includegraphics[width=\columnwidth]{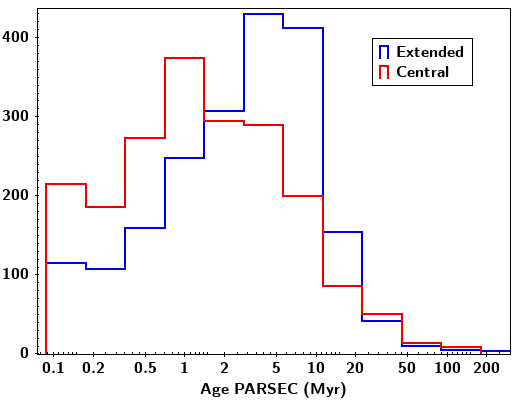}
\caption{Age histograms for the extended sample of the ONC (blue lines). We show the age distributions inferred using MIST (upper panel) and PARSEC (lower panel) models. For comparison, we include the histograms of the central sample normalized to the extended sample size (red lines). The normalization factor is the ratio between the number of stars in the extended sample and the number in the central sample.}
\label{fig:AgeHistogramONC}
\end{figure}

In Fig.~\ref{fig:MassHistogramONC} we now show the resulting mass histograms of the extended (blue) and central (red histograms) obtained with the MIST (top) and PARSEC (bottom) models.  Although the distributions from both models are again statistically different, we will show in \S\ref{subsec:sigmav_ONC} that our result based on the velocity dispersion of the stars per mass bin is robust, regardless the model used.

\begin{figure}
\includegraphics[width=\columnwidth]{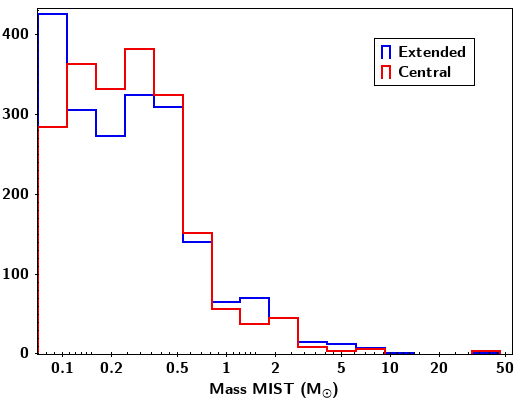}
\includegraphics[width=\columnwidth]{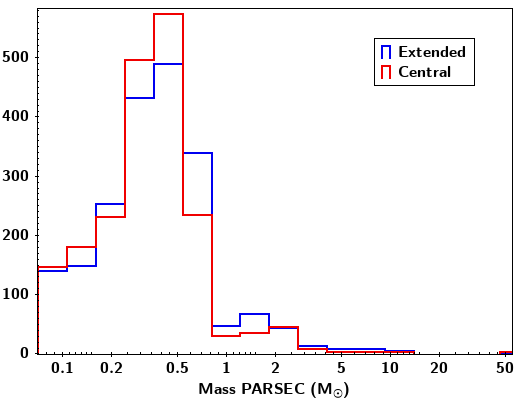}
\caption{Mass histograms of the ONC. Symbols are similar to those in Fig. 5.}
\label{fig:MassHistogramONC}
\end{figure}

}

\subsection{Velocity dispersion per mass bin in numerical simulations. Gravity vs turbulence.} \label{subsec:sigmav_runs}

\begin{figure*}
\includegraphics[width=2.\columnwidth]{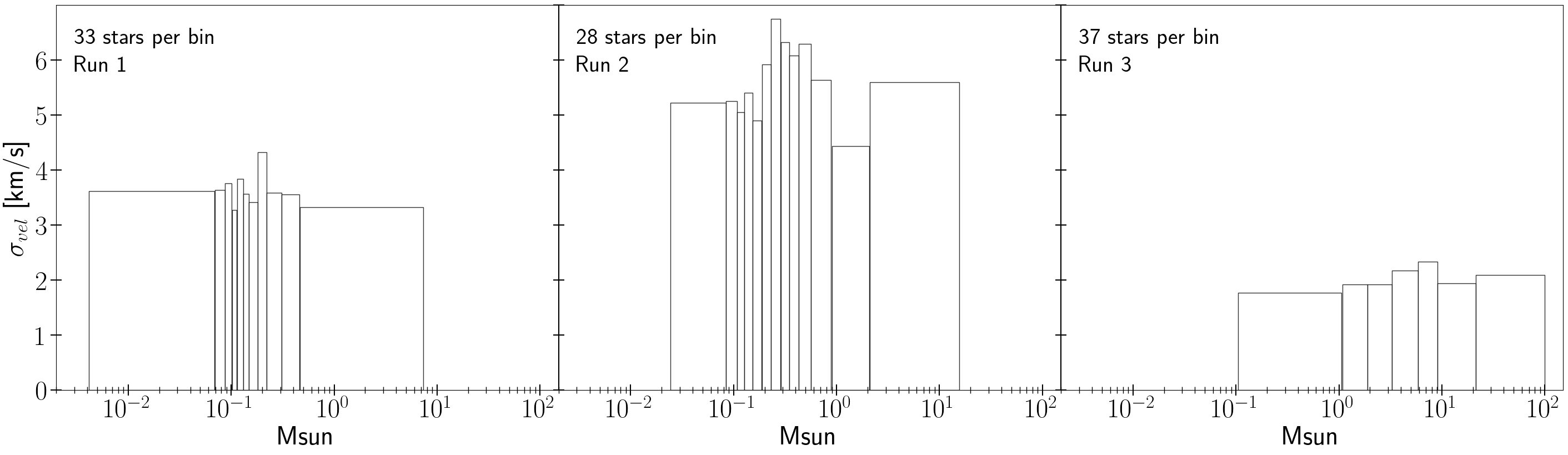} 
%
\caption{Velocity dispersion per mass bin for sinks in runs of global hierarchical and chaotic collapse (runs~1-3). The velocity dispersion, although shows fluctuations, is independent of the mass of the stars, as expected from violent relaxation.}.
\label{fig:sigmav_m:GHC}
\end{figure*}

Fig.~\ref{fig:sigmav_m:GHC} shows the velocity dispersion of the sinks in different mass bins, at $\sim$1$t_{\rm ff}$ for the clusters in our simulations of global, hierarchical and chaotic collapse (runs $1-3$, see \S\ref{sec:methods}). As it can be seen, despite the variety of initial conditions in our suite of simulations, all of them exhibit a fairly flat velocity dispersion as a function of the mass bin of the stars. This behaviour is the expected result for a cloud undergoing a dynamical collapse, and thus, suffering a violent relaxation process \citep{Lynden-Bell67}. 

\begin{figure*}
\includegraphics[width=2.\columnwidth]{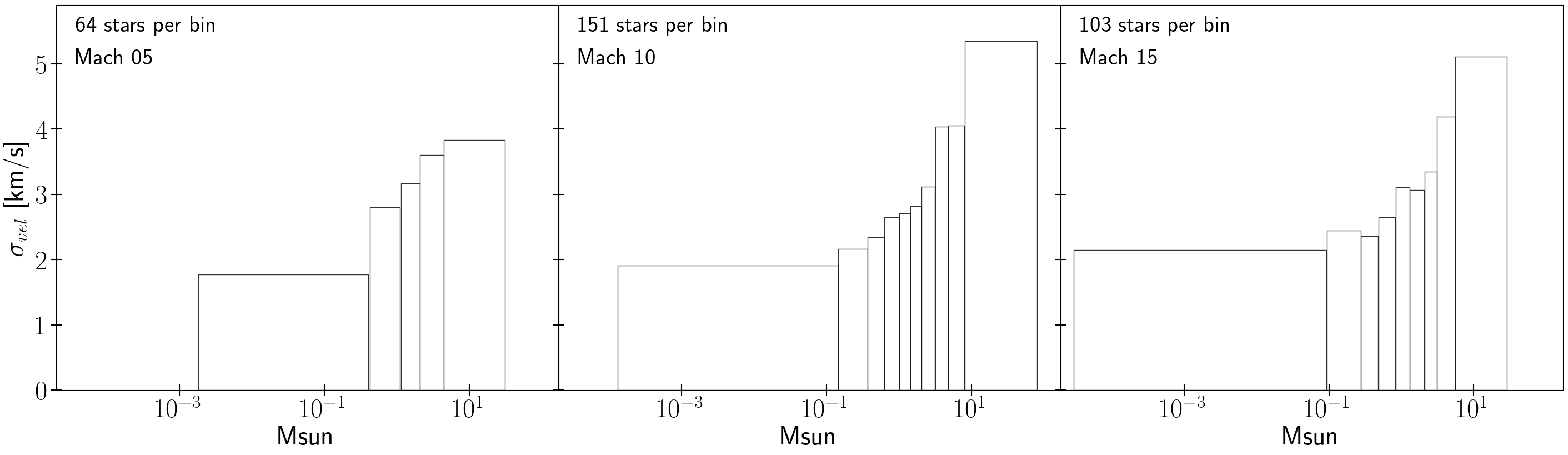} 
\caption{Velocity dispersion per mass bin for sinks in runs of turbulent molecular clouds (runs~5, 6 and 7). The velocity increases for larger masses.  }
\label{fig:sigmav_m:turb}
\end{figure*}

In contrast, the case of sinks in simulations of clouds with forced turbulence is substantially different. In Fig.~\ref{fig:sigmav_m:turb} we plot the velocity dispersion per mass bin of stars formed after stirring our turbulent box during 5 dynamical crossing times, and right after gravity is turned-on and allows the formation of sink particles.  This figure shows that the velocity dispersion increases with the mass of the bin by a factor of $\sim$2 when the mass increases by a factor of 10, suggesting a substantially different process than collapse, which produces `hotter' (in the kinetic sense, i.e., with larger velocity dispersion) massive stars than low-mass stars.

Thus, we can argue that, while collapsing clouds will produce stars that exhibit constant velocity dispersion per mass bin due to the violent relaxation, turbulent environments will produce stars that are kinetically segregated in mass, i.e., with the more massive stars having larger velocity dispersion than the less massive stars.

\subsection{Kinematic evolution of a cluster under formation}\label{subsec:kinematics}

In order to characterise furthermore the kinematics during the formation of the clusters in our simulations, we define the expansion factor as
\begin{equation}
  {\cal{EF}} \equiv\langle\hat{r}_* \cdot \mathbf{v_*} \rangle = \frac{1}{N}\sum_{j=1}^N \big(\hat{r}_{i}\ \ \upsilon_{i}\big)_{j*}
  \label{eq:ExpansionFactor}
\end{equation}
where $\hat{r}_*$ is the unit vector of the star measured from the center of mass of the cluster, $\mathbf{v}_*$ its velocity vector, and $i=1,2,3$ are the Cartesian coordinates, such that the Einstein convention for repeated indexes is adopted. The summation runs over the $N$ stars in the sample. 

\begin{figure}
     \includegraphics[width=\columnwidth]{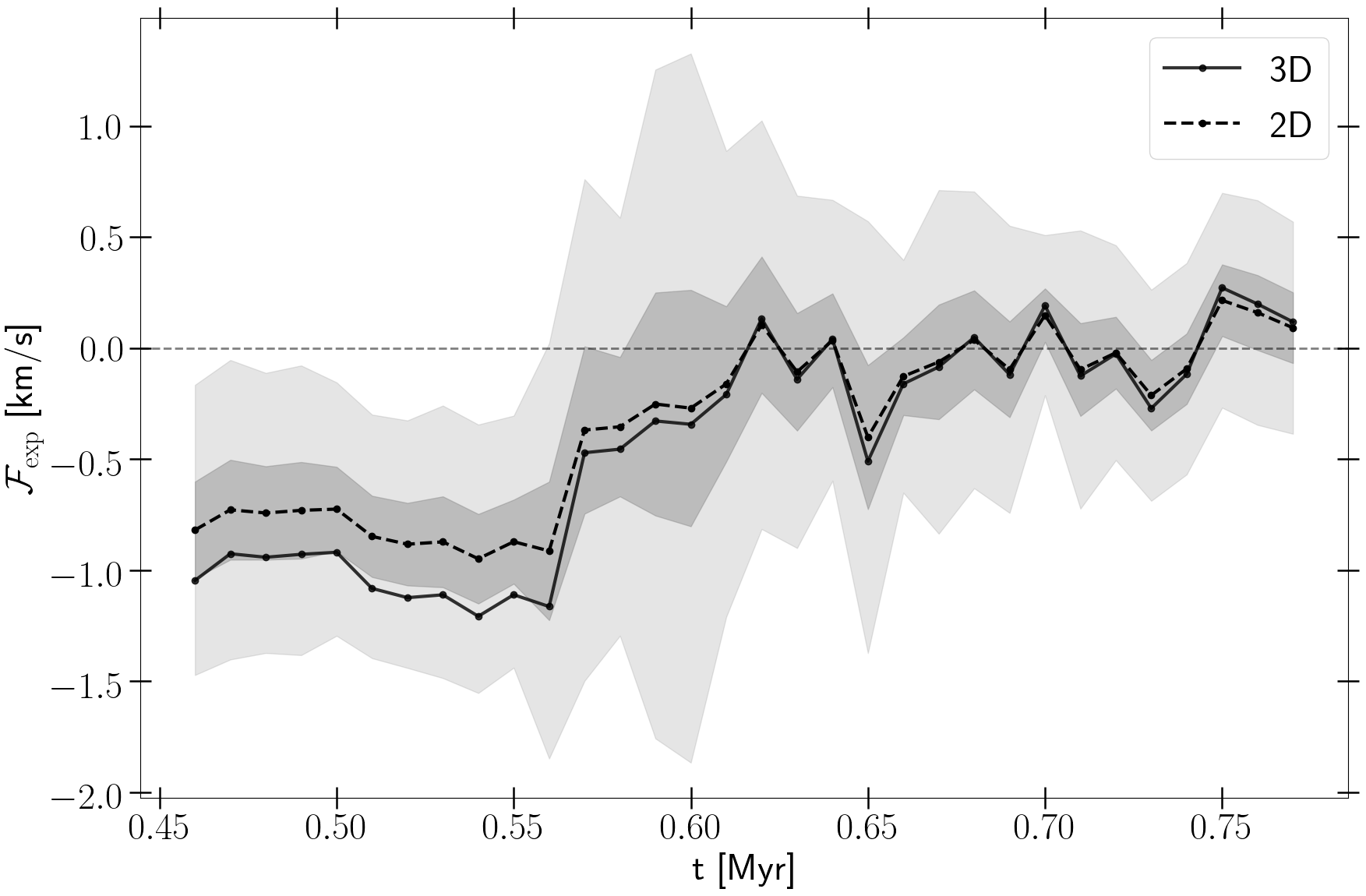}
    \caption{Evolution of the expansion factor ($\cal{EF}$) for the cluster formed in \citet{Kuznetsova+15}.} 
    \label{fig:ExpansionFactorK15}
\end{figure}

In Fig.~\ref{fig:ExpansionFactorK15} we plot the expansion factor ($\cal{EF}$) as a function of time, for the cluster formed in \citet[][a detailed study of the expansion factors in the full set of simulations is presented in Bonilla-Barroso et al., in preparation]{Kuznetsova+15}. The run was stoped at $t\sim$~0.7~Myr, one free-fall time of the gas at its initial density $\rho_0$ in the box,

\begin{equation}
  \tauff = \bigg( \frac{3\pi}{32 G\rho_0} \bigg)^{1/2} \sim 3.4~{\mathrm{Myr}}~ \bigg(\frac{n}{10^2~\mathrm{cm^{-3}}}\bigg)^{-1/2}
  \label{eq:tauff}
\end{equation}
(where $G=6.67\times10^{-8}$ gr cm\alamenos3 s\alamenos2 is the constant of gravity, $n=\rho/\mu m_H$ is the number density, $m_H$ is the mass of the atom of hydrogen, and $\mu=2.36$ the mean molecular weight, assuming solar abundances in MCs). The dashed line in this figure, labeled as 3D, denotes the exact calculation as defined in eq. (\ref{eq:ExpansionFactor}). However, since proper motions only measure a particular projection of the actual 3D data, observations can be skewed by the particular point of view between the earth and the cluster. Thus, we also computed the 2D version of eq. (\ref{eq:ExpansionFactor}) through 1,000 different random directions, and used only the 2D motions projected in a plane perpendicular to each line of sight. As a result, in Fig.~\ref{fig:ExpansionFactorK15} the solid line, labeled as 2D, shows the mean value of the 2D version of the expansion factor (\ref{eq:ExpansionFactor}), while the dark and light shaded areas represent the rms and 3~rms values around the mean, respectively.

Several lessons can be learned from this figure. First of all, it shows us that, at the beginning, the expansion factor appears to be negative, as expected for any collapse process. However, as the potential well increases and more gas and stars fall into the center, the clusters suffer a series of expansions and contractions (we avoid to call them oscillations because that may give the false impression of a stationary system, which is not the case, since the cluster continues accreting gas and stars, and forming additional stars).  This is the period of time during which the velocities are seriously randomized \citep[Fig.~7 by][]{Kuznetsova+15}. 

Naively, one may think that this randomization can give rise to a more dynamically relaxed configuration. However, the relaxation time $\trelax$, defined by eq.~(\ref{eq:trelax}), is still large compared to the evolutionary time. In the present case, the relaxation time (\ref{eq:trelax}) is of the order of $\trelax\sim$12$\tcross$, since $N_*\sim800$, and increasing with time as more stars are born. Thus, assuming that the time span in each oscillation of the cluster is precisely one crossing time $\tcross$, the cluster has not had time to  relax. Thus, even though the velocities are randomized, the velocity dispersion of the stars in a cluster formed in a global collapse process  is independent of the mass of the stars \citep{Lynden-Bell67}, as shown in Fig.~\ref{fig:sigmav_m:GHC}.

A second and important point to mention is that, statistically speaking, even though the cluster has formed during a global collapse of its parental cloud, the detailed value of the $\EF$ depends on the particular 2D projections in which it is observed, as the shaded areas in Fig.~\ref{fig:ExpansionFactorK15} shows, as well as the exact time in which the core is observed. Thus, attributing a particular dynamical state to a cluster that it is still in the process of formation and has substantial amounts of gas in its surroundings, based on the current proper motions \citep[e.g., ][]{Rivera+17, Dzib+18, Ward+18, Ward+20}, should be taken with caution. 


\subsection{Velocity dispersion per mass bin in the ONC}\label{subsec:sigmav_ONC}

\begin{figure*}
\centering
\includegraphics[width=2\columnwidth]{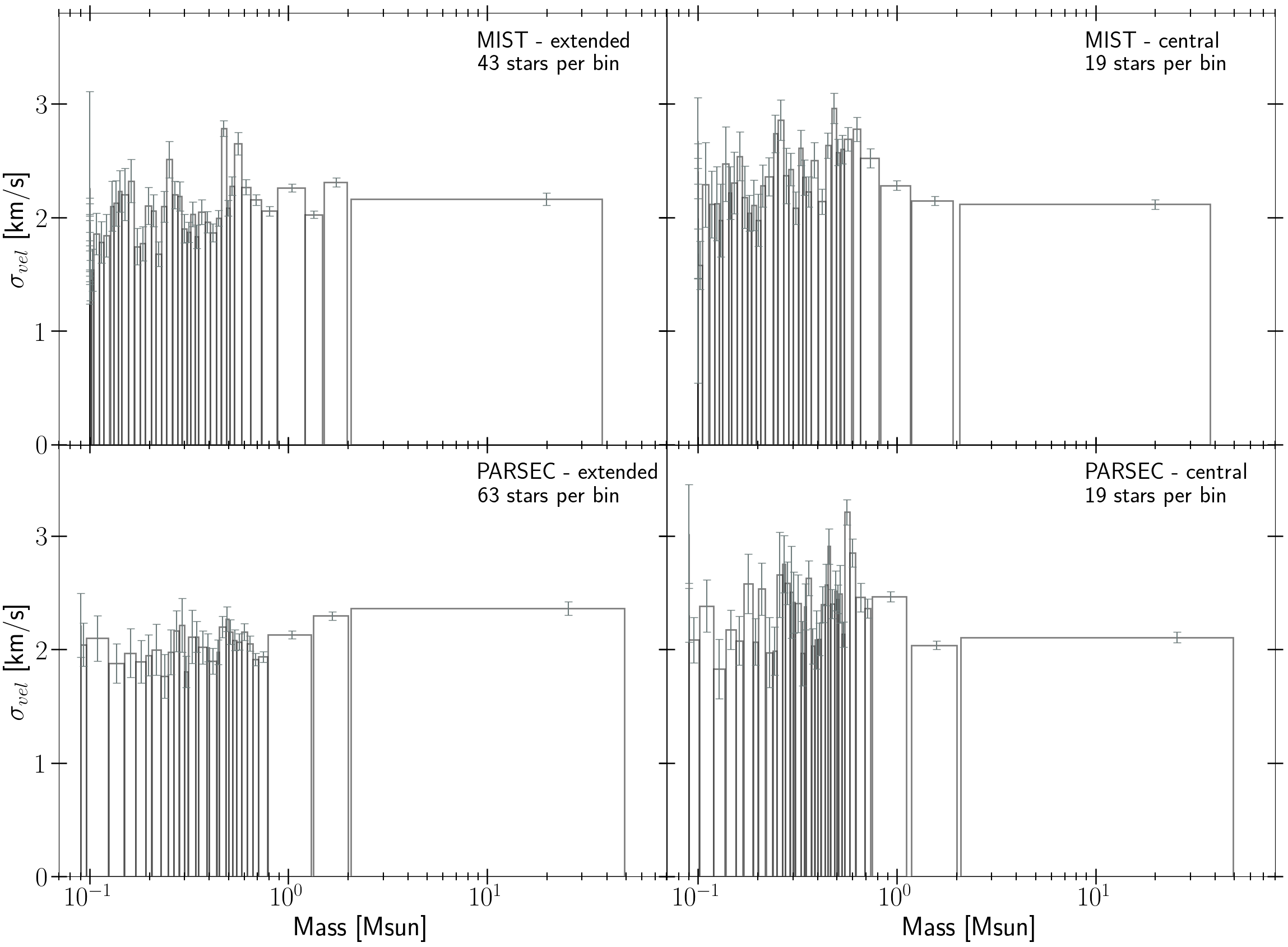} 
\caption{Velocity dispersion per mass bin for stars in the 7~pc size box (left column), and (b) {central 1.5~pc} of the ONC (right column). In the upper panels we used the MIST models in order to infer the masses, while the lower panels we used the PARSEC models. The velocity dispersion per mass bin remains nearly constant in both cases, despite the fluctuations and the models used, indicating that the ONC has been suffering global collapse, rather than been a turbulent region surviving many free-fall times.}
\label{fig:sigmav_m:ONC}
\end{figure*}

In Fig.~\ref{fig:sigmav_m:ONC} we show the velocity dispersion per mass bin of our sample of stars in the ONC. Left panels contain stars within the $\sim$1.5~pc central ONC, while right panels  contains the stars of the more extended ($\sim$7~pc along north-south) region. The masses in the upper panels were computed using the MIST models, while in the lower panels the PARSEC models were used. It is clear from this figure that the velocity dispersion of the stars per mass bin has no dependency with the mass of the bin, a result that spans a factor of $ \gtrsim$ 60 in mass. This result appears to be consistent with the scenario of collapse, and clearly calls into question the somehow spread idea that the ONC formed out of a dense core supported by turbulence against global collapse along $\sim$10 free-fall times at the current density, and forming stars during this time.

\begin{figure}
\includegraphics[width=\columnwidth]{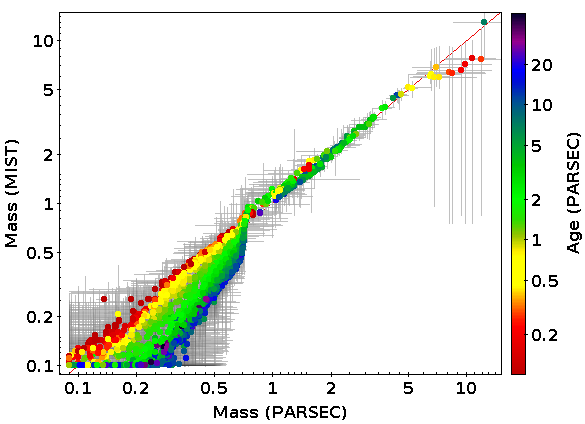}
\caption{Masses of the stars inferred for the MIST models ($y$ axis) {\it vs.} the PARSEC models ($x$ axis). Notice the clear, one-to-one correlation of the massive stars, indicating that the most massive bins computed in \S\ref{subsec:sigmav_ONC} contain the same stars, and thus, their velocity dispersion are essentially the same.}
\label{fig:massMist_vs_massPARSEC}
\end{figure}

An important question to address is in order. If the MIST and PARSEC models produce statistically different mass distributions (as shown in Fig.~\ref{fig:MassHistogramONC}), could the none-dependency with mass of the velocity dispersion (Fig.~\ref{fig:sigmav_m:ONC}) be the result of actual masses of the stars 
mixed up in different bins, such that actual differences in the velocity dispersion of the stars with different masses is averaged out?  The answer is no. As shown in Fig.~\ref{fig:massMist_vs_massPARSEC}, it is clear that the massive stars in both models are not mixed with the low mass stars, and thus, the more massive bins (above $\sim$1~\Msun) contain the same stars in both models, and thus, the same statistics. Any increase or decrease of the velocity dispersion, at least of the most massive stars, will be detected, regardless the model used.

\section{Discussion}\label{sec:discussion}

The formation and evolution of young stellar clusters has been matter of debate for long time. Many works have focused on the spatial structure of young clusters. Whether there is (or not) {\it spatial-}mass segregation, whether such segregation is primordial or the result of a relaxation process, or what is the density structure of the stars, are just a few questions that have been addressed by a number of authors \citep[e.g., ][]{Hillenbrand_Hartmann98, Goodwin_Whitworth04, Goodwin+07, Allison+09, Moeckel_Bonnell09, Bate12, Agertz+13,Parker+16, Alcock_Parker19}. 

Less attention has been put to the velocity structure of the stars in clusters in general, and to the {\it velocity dispersion-}mass segregation in particular. For the former, some works have discussed whether OB associations are expanding, contracting, rotating, or randomized \citep[e.g.,][Bonilla-Barroso et al. 2021, in preparation]{Rivera+17, Dzib+18, Roman-Zuniga+19, Ward+20}, while others have tried to address whether clusters are in  equipartition, or develop the \citet{Spitzer69} instability. In this sense, \citet{Wright_Parker19} have analyzed the  {\it velocity dispersion-}mass segregation in the Lagoon Nebula open cluster, finding that the massive stars exhibit larger velocity dispersion than low-mass stars, just as our simulations of turbulence environments (Fig.~\ref{fig:sigmav_m:turb}). 

More than 50 years ago, trying to understand the process of galaxy formation, \citet{Lynden-Bell67} discussed the relaxation that occurs in a collapsing stellar system, where the mean gravitational field is not in steady state, but undergo important variations within a free-fall time scale. \citet{Lynden-Bell67} showed that, due to the changes in the gravitational potential, the individual stellar energies are not conserved, but instead, the gain or loss of energy per unit mass by any star does not depend on its mass. In other words, the velocity dispersion of the stars, per mass bin should be constant when a process of collapse is occurring.  While the systems devised by \citet{Lynden-Bell67} were galaxies, the physics is applicable to young stellar clusters, though. Under this idea, one can expect that if young clusters are formed in a process of collapse within a few free-fall timescale, they should exhibit a velocity dispersion that is independent of the mass of the stars. 

The standard picture of cluster formation, nonetheless, suggests that these form in massive clumps where turbulence is the main physical agent providing support against collapse over several free-fall times \citep[e.g.,][]{Krumholz_Tan07, Krumholz+19, Ward+20}. 
In such environment, massive stars are born in massive cores which in turn, are formed by stronger turbulent shocks \citep[e.g.,][]{McKee_Tan03, Padoan+16}. 
Thus, one can expect that massive cores will have a larger velocity dispersion as a result of larger velocity shocks in random directions, while low-mass cores require substantially smaller Mach number shocks to be formed, and thus, their velocity dispersion is expected to be smaller. 

From the numerical simulations, the statistics presented in the previous section shows that looking at the {velocity dispersion per mass-bin} could be a good way to discriminate between possible star formation scenarios occurring in star forming regions. On one hand, turbulence-induced star formation tend to produce strong {\it kinetic-}mass segregation in the stellar population: massive stars exhibit larger velocity dispersion than low-mass stars. On the other hand, gravity-dominated star formation tends to produce no {\it kinetic-}mass segregation, but stars exhibit a flat (within the statistical fluctuations) velocity dispersion per bin mass.

In addition, we also have shown that the ONC exhibits no signs of {\it velocity-}mass segregation, strongly suggesting that it has been formed by a process of global collapse of its parental clump, within one or two free-fall times at its initial density. This result is valid not only for the central, compact ONC, but for the more extended, 7~pc size region.

It is important to recall that the ONC and Orion-A exhibit also a variety of observational features that are also present in numerical simulations of global collapse, as has been shown by \citet{Kuznetsova+15} and \citet{Kuznetsova+18}. In particular, (a) \citet{Hillenbrand_Hartmann98} found that the ONC is  {\it spatially-}mass segregated, (b) the velocity dispersion of the stars and gas increases towards the sites of collapse \citep{Hacar+17b, DaRio+17}; or (c) that the stars in the ONC do exhibit random proper motions \citep[][]{Kounkel+18}. 

The proposal that the ONC has been formed within a few free-fall times at its initial density is at odds with the idea that the ONC has survived by  $\sim$10 current (high-density) free-fall timescales, forming stars at a low pace, as has been suggested by a number of works \citep[e.g., ][]{Tan+06, Krumholz_Tan07, Krumholz+19}. In what follows, we discuss the arguments that have been posed against the collapse model and/or favouring the turbulent model, showing that, as a matter of fact, the data related to the ONC is compatible with the collapse model.


\subsection{Proper motions, rate of expansion, and virial balance.}\label{subsec:ProperMotions}

A frequent misconception is that the proper motions of the stars in collapsing clouds are necessarily inward motions, if the observation occurs during the process of contraction of the cloud, and outwards, once the remaining gas cloud has been removed by stellar feedback and the potential well disappears \citep[e.g., ][]{Ward+18,Ward+20}.
In reality, the hierarchical and chaotic collapse that occurs in molecular clouds is never as simple as such picture. In fact, as collapse proceeds, denser regions keep incorporating more gas and stars from the surroundings, increasing the potential well, and allowing the stars in the cluster to rapidly randomise their velocity vectors as the potential well increases as the result of the infall of material \citep[see Figs.~7 and 13 from ][]{Kuznetsova+15}. 

Several authors have argued  that the ONC should be clearly expanding if it was formed by the GHCC model, because the gas has already been removed by the stars \citep[e.g.,][]{Krumholz+19}. We want to stress several points in this regard. First of all, there is a considerable amount of molecular gas that remains behind the optically visible cluster \citep[e.g., ][]{ODell01}. As pointed out by \citet{Hillenbrand_Hartmann98}, a reasonable estimate is $\sim$~4000~\Msun\ of gas within a diameter of 4 pc \citep[2200~\Msun within 2 pc, see][]{Bally+87}.

In addition, as we have shown in \S\ref{subsec:kinematics}, during a substantial fraction of the formation timescale of a still embedded cluster, the expansion factor $\EF$ (eq. [\ref{eq:ExpansionFactor}]) can oscillate around zero, even if the cluster has been formed by the global collapse of its parental clump\footnote{We recall also that the detailed value of $\EF$ could depend upon the exact time in which it is observed, and its 2D calculation from proper motions, furthermore, may depend on the detailed projection it has, as discussed in \S\ref{subsec:sigmav_runs}.}. Furthermore, it is well known that there is a population of embedded stars in the region, and thus, the cluster can still be under construction, with stars and gas falling in, as it occurs in numerical simulations \citep[e.g., ][]{Kuznetsova+15}. 

Finally, it is important to recall that the original problem posed by \citet{Krumholz+19} is not exclusive of the GHCC model. In other words, the fact that ONC exhibits a velocity dispersion that exceeds its virial value (and thus, must be expanding), but has a small expansion factor, is a problem that can be posed to any model of molecular cloud evolution and cluster formation, regardless of whether the original cloud is collapsing, or supported by turbulence.



We conclude, thus, that neither the apparent randomization of velocities in the ONC, or its small rate of expansion,
can be an argument to dismiss the global collapse of a larger clump as progenitor of the ONC. The fact that the ONC region still has substantial amounts of gas it its background should be considered  when arguing whether expansion should be observed or not.

\subsection{Age spreads, stellar distribution and free-fall times in the ONC.}\label{sec:AgeSpreadONC}

Its been argued that the stars in the ONC exhibit an age histogram with a peak at $\sim$6--8~Myr \cite[][see panels c and d in their Fig.~13]{Krumholz+19}, for an area similar to our extended region, and nearly constant for the central cluster. These authors argue that the ONC has been forming stars over the last 6~$-$~8~Myr, and thus, over $ \gtrsim$10~$\tauff$.  Several points need to be addressed in this respect. 

\subsubsection{Age spreads}

First of all, as the ages presented by \citet{Krumholz+19} were taken from \citet{Kounkel+18}, they have been computed using the PARSEC models.  Our ages estimations using the PARSEC models show, indeed, a quite similar distribution, though the peak of our distribution is still at $ \lesssim$5 Myr\footnote{It is worth mentioning, furthermore, that in our calculations we used Gaia EDR3, as well as individual values of the extinction through the stars, while the estimations by \citet{Krumholz+19} used Gaia DR2, and a single extinction law.} (see blue histogram in the lower panel of Fig.~\ref{fig:AgeHistogramONC}), not at 6--8 Myr  \citep[as shown in panel c of Fig.~13 from][]{Krumholz+19}. Moreover, the stellar parameters in \citet{Kounkel+18} could have unphysical systematics, likely due to mismatches
between the empirical and theoretical spectra \citep{Olney+20}, and this can affect the age distribution presented by \citet{Krumholz+19}.

Nonetheless, our estimations using the MIST models (blue histogram in the upper panel Fig.~\ref{fig:AgeHistogramONC}) show clearly younger ages. The peak at $t\sim$2~Myr, respectively clearly reduce the star forming timescale. Whether one models or the others reproduce better the evolutionary tracks of pre-main sequence stars, and thus allow us to compute better statistics of the ages of newborn clusters has no straightforward solution.  For that purpose, it will be necessary to have better evolutionary models, as well as more detailed photometry of the stars in the ONC.  In any event, the relevant point that we want to raise is that the conclusions posed by \citet{Krumholz+19} in the sense that the ONC has lived many free-fall times will not necessarily follow.  

\subsubsection{Distribution of stars in the ONC}

\begin{figure}
 \includegraphics[width=\columnwidth]{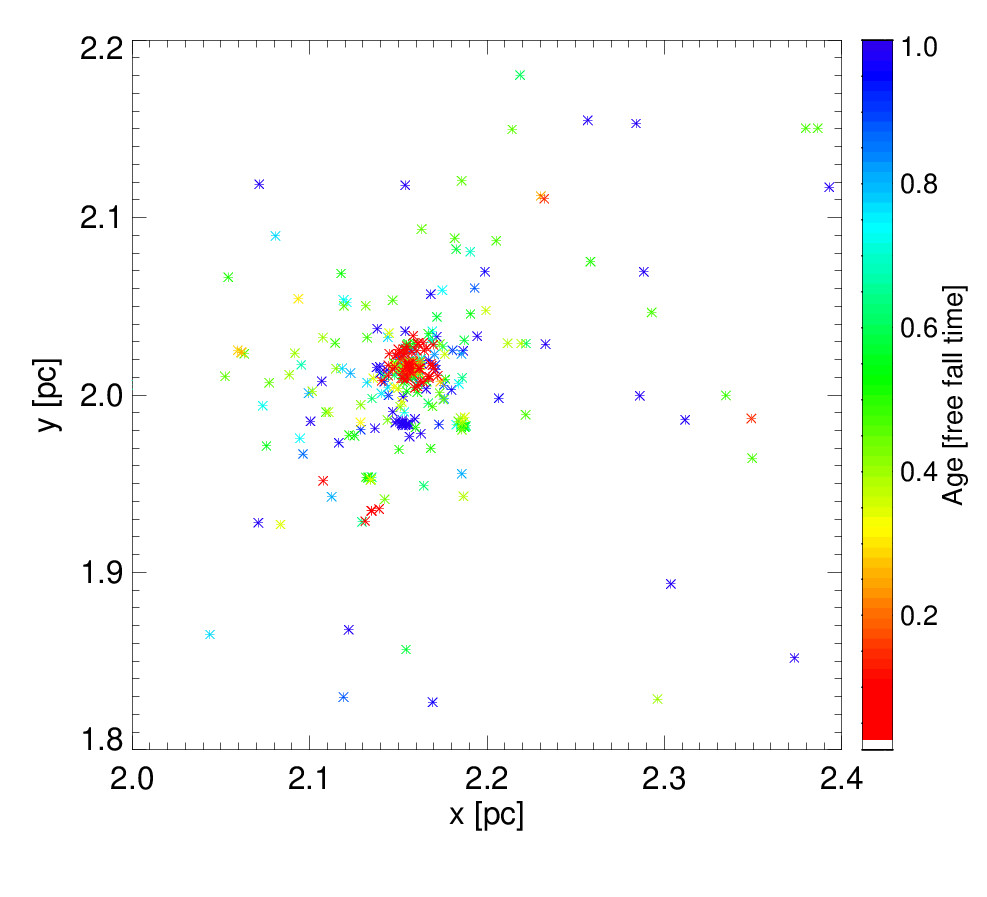}
  \caption{Age segregation in the main cluster of \citet{Kuznetsova+15}. As can be seen, younger stars are concentrated towards the central region of the cluster, while the older stars are spread over larger areas. }
    \label{fig:SpatialAgeSegregation}
\end{figure}

The argument of the ONC being a core supported by turbulence over many free-fall times utterly fails to account for the change in spatial distribution of stars of differing age. Specifically, the stars in the wider 7 pc region of the ONC have average ages $\sim 5$~Myr, while the members of the 1.5 pc region, much more spatially concentrated, have characteristic ages closer to 1--2 Myr  (note that while the magnitudes of the ages depend upon which evolutionary tracks are used, the {\it relative} age differences persist, regardless the tracks used). Furthermore, the protostars, i.e. the very youngest population, are even more spatially confined, with a large fraction of them concentrated in the dense, narrow ``integral shaped filament'' (see Figure 14 in  \cite{Megeath+2012}).  This sequence of decreasing age with decreasing spatial scale is not consistent with a relatively unevolving model where turbulent support prevents collapse, let alone one which maintains support for many free-fall timescales. On the contrary, this is exactly what one would expect from a collapse model, since the first stars are formed in a larger, more spread area, while the younger stars necessarily are concentrated towards the more collapsed regions, as can be seen in Fig.~\ref{fig:SpatialAgeSegregation}, where we show the positions of the sink particles in run~2, colored by their age, with red asterisks denoting the younger sinks, and blue asterisks denoting the older ones, in units of the initial free-fall time of the simulation.

\subsubsection{The free-fall time of the ONC}\label{sec:FreeFallTime}

It is frequently argued that the free-fall timescales of MCs are substantially shorter than the life timescales of the molecular gas, and thus, that the star formation spans several free-fall timescales producing stars at a low star formation rate per free-fall time \citep[$\SFRff$, see][]{Tan+06, Krumholz_Tan07, Evans+21}. In particular, for the ONC, \citet{Krumholz+19} argue that the free-fall timescale of the gas in the central cluster is of the order of $\sim$0.6~Myr, and thus, a histogram exhibiting stellar ages with a peak at $\sim$6--8~Myr (extended region), 
suggest that the ONC has been forming stars over at least 10 free-fall timescales. Several flaws can be distinguished. 

First of all, the age estimations of the stars in the central core are substantially smaller, as discussed in the previous section and seen in Fig.~\ref{fig:AgeHistogramONC}.

But more important, the estimation of the free-fall timescale of the ONC is based on the current volumetric density of the molecular gas. As pointed out by \citet{Vazquez-Semadeni+19}, however, this value severely underestimates the time lapse that a larger and less dense cloud has to spend in order to achieve its present state. In fact, the gas that produced the extended region of the ONC \citep[Ori A in][]{Krumholz+19} could have had very well an original density of $n\sim$ some $\times$~10$^2$~cm\alamenos3, and collapsed in a few Myr, as given by eq.~(\ref{eq:tauff}), and producing, thus a characteristic age histogram with a peak around $\sim$2.3~Myr. Furthermore, the central core of the ONC may very well have had larger densities, of the order of $\sim$10$^3$cm\alamenos3, and thus, have collapsed in the last $\lesssim$1~Myr. 

In a globally collapsing cloud, thus, the current free-fall timescale based on the current density is, by construction, substantially smaller than the age spread of the stars, which have been forming since earlier times, and thus, such comparison cannot in any way discard or favour one model of star formation over the other. 

 \subsubsection{Dynamically older stars at the ONC?}
 
 Leaving aside the question of which free-fall time is relevant in a region where densities range over orders of magnitude, we still can play the game of trying to understand how dynamically old are the stars in each region.  \citet[][]{Krumholz+19} argue that the central ONC, being younger, is dynamically older than the stars in the external region. A comparison between their assumed free-fall times and the ages shown by the models however, shows that this is not the case. The central region has stars with ages of the order of 1-2~Myr. Assuming that its free-fall time is $\sim0.6$~Myr, as proposed by these authors, its stars are 2-3 free-fall times old.  On the other hand, the external region has ages of $\sim$6~Myr, but its free-fall time is $\sim$2--3~Myr, and thus, in either case, both regions have dynamically the same age. 
 
 It should be stressed that, in terms of their dynamical processes, in a violent relaxation process the timescale that matters is the timescale of the variation of the gravitational potential, which necessarily is the free-fall time. Thus, the older stars in the cluster are also dynamically older stars.

\subsection{The efficiency of star formation in the ONC}

There are several similar definitions of the efficiency of star formation, but the most simple one is the amount of gas of a cloud that has been converted into stars. In clusters where stellar feedback has cleared up at least partially the mass of the parent cloud, it becomes necessary to estimate the total mass that was involved in the formation of the ONC in the first place. \citet{Krumholz_Tan07} argued that between 6,700 and 15,000~\Msun\ were involved in the original cloud that gave rise to the ONC, and used the value of 4,500~\Msun\ quoted by \citet{Hillenbrand_Hartmann98} for the mass in stars. With these numbers one can compute a ``present-day'' efficiency between 0.3 as a minimum and 0.67 as a maximum.

We first stress that these numbers are independent on the assumed scenario of star formation. In other words, whether it occurred within one or many free-fall times, if the mass of the original cloud and of the stars are the quoted above, the present day efficiency is between 0.3 and 0.67. 

It has been argued, furthermore, that while the turbulent scenario may reach such final efficiencies along many free-fall times, at a low efficiency per free-fall time, the collapsing scenario would require ``extreme efficiencies'' integrated over only one {\it current}, or {\it present-day} free-fall time (of the order of a few $10^5$~yr), of the order of $\epsilon>0.3$, in order to produce bound systems \citep{Krumholz+19}.     
There are several problems with this argument. The main one, as in the previous case, it assumes that the star formation occurs within one {\it present-day} free-fall time, i.e., the free-fall time estimated at the current mean gas density of the cluster. As commented out in the previous section, this time is not the relevant free-fall time involved in the  hierarchical and chaotic collapse scenario, as we discussed in the previous section, but the free-fall time from the beginning of the cloud contraction, when the cloud was substantially less dense. 

In addition, it should be recalled that the aforequoted 4,500\Msun\ in stars in the ONC is estimated from virial equilibrium, and, according to \citet{Hillenbrand_Hartmann98}, $\sim$50\%\ of that mass comes from the gas in the cloud behind the optical cluster. That reduces a factor of 2 the apparently ``extreme'' efficiency that the collapse model will require.  But even more, recent estimations from \citet{DaRio+12} quote a total of 1,000~\Msun\ in stars, putting the efficiency of star formation for the ONC below 10\% if the original mass of the cloud was indeed 15,000~\Msun, as quoted by \citet{Krumholz_Tan07}. Such efficiency could be easily achieved within one {\it original} (few Myr) free-fall timescale, without implying a ``huge'' star formation efficiency.

There is also the concept of ``efficiency per free-fall time'', i.e., the efficiency that would have occurred after one free-fall time. Obviously, this is different from the terminal efficiency if the process of star formation has taken more than one free-fall time. Turbulence models assume that clusters are formed along $\sim$10 or more free-fall timescales, while hierarchical and chaotic collapsing models collapse within $\sim$1$-$2~free-fall timescales.  Thus, the estimated efficiency per free-fall time is of the order $\sim$0.005$-$0.01 for the turbulent models, and between $0.02-0.8$ for the global collapsing model. But it should be noticed that the free-fall timescale is substantially different in each model, of the order of few $10^5$yr for the turbulent model, and of the order of few Myr for the hierachical and chaotic model, as we discussed in \S\ref{sec:FreeFallTime}.  The very concept of efficiency per free-fall time somehow has not much relevance if the free-fall timescale is not constant, but varies with time, as occurs in the collapse models.  In these, in fact, the star formation becomes accelerated as collapse proceeds, a feature that is present in star forming regions  \citep{Hartmann+12}, and that models with constant efficiency per free-fall time cannot reproduce. 

Finally, it should be noticed that, part of the observational evidence that the ``small'' and almost independent on the density $\SFRff$ is based on the assumption that the HCN traces dense gas, of the order of $6\times10^4$cm\alamenos3. However, \citet{Kauffmann+17} has shown that HCN actually traces gas with densities of the order of $n\sim$~900~cm\alamenos3, substantially less dense than it has been thought. In fact, if such density had been  used in Fig.~5 from \citet{Krumholz_Tan07}, it will had been difficult to argue that the $\SFRff$ was small and independent on the density, as proposed by these authors.

%

\subsection{Protocluster vs cluster mass densities }

Another wrongly posed argument against the GHCC model is that clusters have to pass through a phase in which their mass density should be larger than the mass density of the resulting stellar clusters \citep[][see arguments regarding their Fig.~14]{Krumholz+19}. 

First of all, it should be noticed that the same problem can be posed to the turbulent model: if a volume of molecular gas forms a stellar cluster, the gas mass density of the parent cloud should be similar or larger to the stellar mass density after cluster formation and cloud dispersal, contrary to what it is shown in \citet{Krumholz+19}'s Fig.~14. 

\begin{figure}
     \includegraphics[width=\columnwidth]{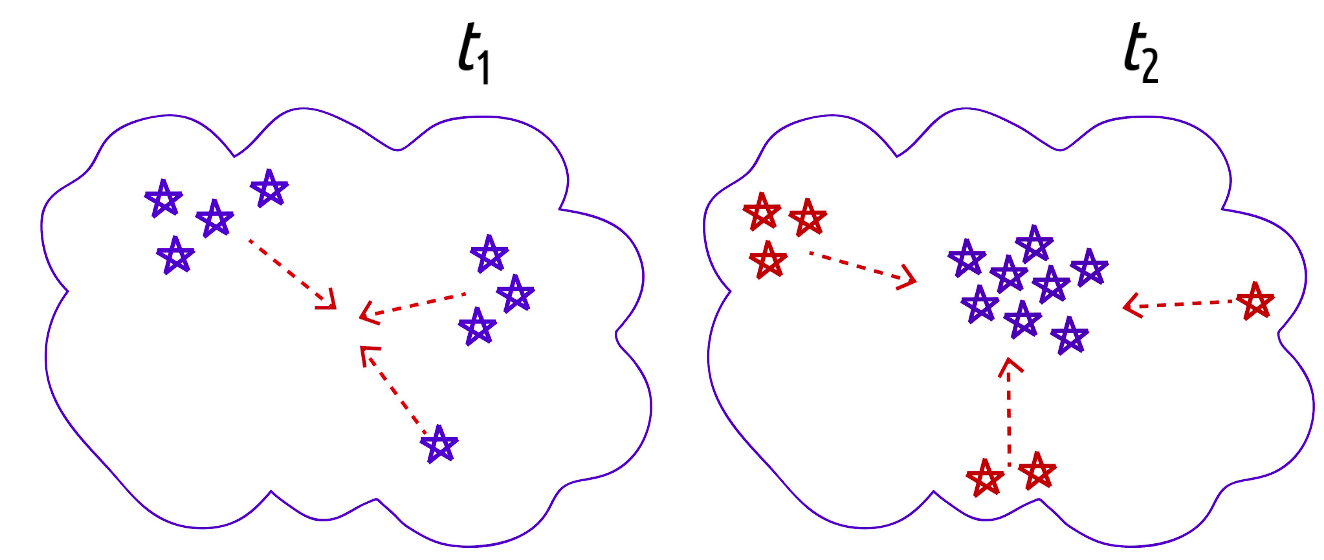}
  \caption{Schematic diagram showing how a stellar cluster is built up in the GHCC model without the need of passing through a phase in which the collapsing core is more dense than the resulting cluster.  In the GHCC, stars are born in the very place of the cluster, as well as in its vicinity \citep[see also Figs. 1 and 2 in][]{Kuznetsova+15}.}
    \label{fig:esquema_starvation}
\end{figure}

Second, this misconception probably arises because clouds are historically thought to be static. Then, if a cluster forms in the densest clump, and the final efficiency is smaller than 100\%, by construction the parent clump should be more dense than the cluster it formed. But as we have argued, hierarchical and chaotic collapse is not as simple as such a picture. On the contrary, numerical simulations of collapsing clouds forming  clusters actually explains with relatively simplicity the observations described by \citet[][]{Krumholz+19}. In these models, clusters form by a combination of stars forming in the central collapsing region, but also by incorporating newborn stars from the vicinity \citep[see Figs. 1 and 2 in ][]{Kuznetsova+15}. During this process, the stellar mass density can become substantially larger than the gas mass density by a combination of the inclusion of newborn stars from the vicinity, but also by gas starvation in the central region, as shown in a variety of numerical simulations \citep[e.g., ][]{Girichidis+12b, Ballesteros-Paredes+15}. This process is also schematically depicted in Fig.~\ref{fig:esquema_starvation}.

\begin{figure}
    \includegraphics[width=\columnwidth]{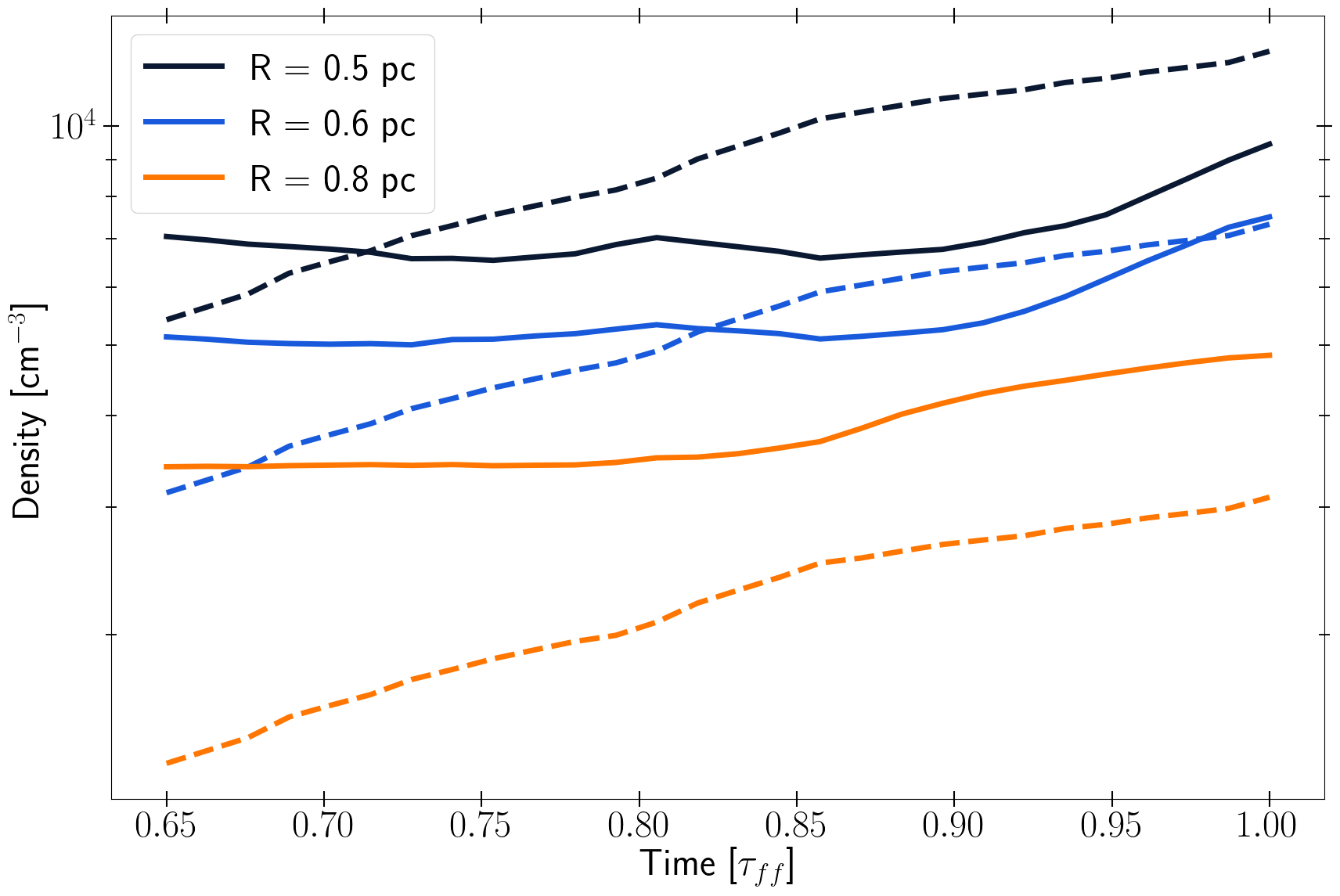}
    \caption{Mass density of gas (solid lines) and stars (dotted lines) of the main cluster by \citet[][]{Kuznetsova+15}, computed on spheres of radii 0.5, 0.6, and 0.8 pc from the center of the cluster. The continuous growing in gas and stars allows that, for certain radii and certain time the mass density in stars exceeds the mass density in gas.}
    \label{fig:density_cluster_core}
\end{figure}

To show that this is the case, we again look at the simulations by \citet[][similar results are found for the other runs]{Kuznetsova+15}, where the stellar clusters are build up during the process of collapse, as gas and stars are incorporated continuously into the central cluster.  In Fig.~\ref{fig:density_cluster_core} we plot the mass density of gas (solid lines) and the mass density of stars (dotted lines) as a function of time, for the last 1/3 of the evolution time (recall that we evolve that simulation over one initial free-fall time). The mass densities are computed over spherical volumes of radii as indicated in the figure, and, at every time, they are centred at the position of the main cluster. 

As it can be seen from Fig.~\ref{fig:density_cluster_core}, during the process of the cluster formation, there is a moment in which the stellar density (dotted lines) becomes larger than the gas density (solid lines).  Only for large enough spheres (purple line), the gas density are always larger than the stellar densities, although it is clear that even in this case, the stellar density approaches to the gas density towards the end of the computation. 

Finally, in addition to what we have discussed, it should be noticed that during the process of collapse, stellar clusters tend to develop strong density concentrations, as the entropy of gravity dominated cluster is not bound: self-gravitating systems have no maximum entropy configuration, and the more concentrated they are, the higher their entropy \citep[][see also Binney \ Tremaine, \S7.3.2]{Lynden-Bell_Wood68}.

Thus, it cannot be surprising that there are no molecular clouds as dense as stellar clusters, and finding that stellar clusters do exhibit larger densities compared to the densities of molecular clouds cannot, in any way, be proof that clusters cannot be formed by a process of global collapse.  In fact, rather than a problem, its a natural outcome from the GHCC model. 

\subsection{Inverse velocity dispersion in the Lagoon Nebula}

\textcolor{black}{\citet{Wright_Parker19} found evidence for an increasing velocity dispersion with mass in NGC 6530, a young ($\sim$2~Myr) stellar cluster at a distance of $\sim$1.3~kpc. They interpret this trend in velocity dispersion as a result of collapse over time to a more compact configuration where the  Spitzer instability causes the massive stars to form a smaller system with a higher velocity dispersion. Wright \& Parker refer to N-body simulations analyzed in \citet{Parker_Wright16} to argue that the observed mass-dependent velocity dispersion requires not only cool collapse but an initial highly substructured spatial distribution of stars.}

\textcolor{black}{These observational results for NGC 6530 differ from ours for the ONC/Orion region, where we find no real evidence for a trend of velocity dispersion with  stellar mass, even though we similarly argue for cluster collapse from subvirial conditions. In considering possible explanations for the difference, it is important to recognize that the velocity dispersions of both low- and high-mass stars will be strongly determined by the depth of the gravitational potential well.  In the case of NGC 6530, the high mass stars tend to be much more spatially concentrated than the low-mass sample (see Figure 1 of \citet{Wright_Parker19}) and so consistent with higher velocity dispersions correlating with deeper gravitational potential. Perhaps more importantly,  \citet{Wright2019} found evidence that on large scales the cluster is expanding, which is clearly inconsistent with a simple model of sub-virial cluster collapse. This possible expansion could be a result of expulsion of most of the initial gas in the region, as suggested by the relatively low extinction to many cluster members and the morphology of molecular gas in the region \citep{Tothill2002}, which mostly lies around the periphery of the cluster.  Both of these apparent features - expansion and lack of dynamically significant molecular gas - differ qualitatively from that of our Orion/ONC sample.}

\textcolor{black}{With regard to differences between the numerical simulations, those of \citet{Parker_Wright16} are pure N body simulations without gas and do not involve collapse from initially much larger scales.  In addition, the simulations of \citet{Parker_Wright16} that show substantial growth of the velocity dispersion of the massive stars assume strongly subvirial initial conditions ($\alpha_{vir} = 0.3$) and/or strong initial substructuring ($\alpha_{vir} = 0.5$, fractal dimension $D = 1.6$); cases of less substructuring for $\alpha_{vir} = 0.5$ show little or no velocity dispersion growth. It is not obvious that our cluster simulations show similar substructuring, and this again may be affected by the presence of dynamically-important gas.}

\section{Conclusions}

In the present work we have analysed numerical simulations of molecular clouds in order to understand the kinematical properties of young stellar clusters. We have found that, as proposed by \citet{Lynden-Bell67}, collapsing star-forming clouds produce clusters whose stars exhibit constant velocity dispersion as a function of mass, while, turbulence-supported clouds exhibit an {\it inverse} mass segregated velocity dispersion, where massive stars have a larger velocity dispersion than low-mass stars. We also show that collapsing clouds exhibit spatial segregation, with older stars been more spread out than younger stars. 

We also showed that both characteristics found in collapsing models are present in the ONC, which exhibits a constant velocity dispersion as a function of mass, as well as spatial segregation \citep[see also][]{Getman+14, Getman+18}, suggesting that it has been formed by a process of global collapse within one free-fall time of its parental cloud. 

In addition, we have discussed several of the criticisms that have been posed to the model of collapse, showing that, frequently, these are more the result of missunderstanding what an actual cloud will do during its collapse.  In particular, we showed that collapsing, cluster forming clouds do exhibit

\begin{enumerate}
  \item Random motions of its stars, 
  \item Small, expansion factors of the ONC and simulated clusters, specially if there is still substantial amount of gas, as it seems to be the case of the ONC,
  \item Spatial stellar age segregation,
  \item Total star formation efficiencies of $\lesssim$~10\%, \label{item:efficiency}
  \item Mass densities larger than those of their parental cloud,
\end{enumerate}
With the exception of point \ref{item:efficiency}, which requires stellar feedback, all of them are the natural outcome of collapsing cloud dynamics. 

\section*{Acknowledgments}

The numerical simulations performed in this work were performed either in the Miztli cluster at DGTIC-UNAM through proposal LANCAD-UNAM-DGTIC-188, and in the Mouruka cluster at IRyA, provided by CONACYT through grant number {\tt INFR-2015-01-252629.}
A.B.-B. acknowledges scholarship by CONACyT. 
J.B.-P.  acknowledges UNAM-DGAPA-PAPIIT support through grant number {\tt IN-111-219}, CONACYT, through grant number {\tt 86372}, and to the Paris-Saclay University’s Institute Pascal for the invitation to ‘The Self-Organized Star Formation Process’ meeting, in which invaluable discussions with the participants lead to the development of the idea behind this work. 
J.H. acknowledges support from CONACyT project No. 86372 and the UNAM-DGAPA-PAPIIT project IA102921.
Support for A.K. was provided by NASA through the NASA Hubble Fellowship grant \#HST-HF2-51463.001-A awarded by the Space Telescope Science Institute, which is operated by the Association of Universities for Research in Astronomy, Incorporated, under NASA contract NAS5-26555.
V.C. acknowledges support from CONACyT grant number {\tt A1-S-54450} to Abraham Luna Castellanos (INAOE).

This work has made extensive use of SAO-NASA Astrophysical Data System (ADS).

\section*{Data availability}
The data underlying this article will be shared on reasonable request to the corresponding author.


\bsp	
\label{lastpage}
\end{document}